\documentstyle[12pt,epsf,epsfig]{article}
\newcommand{\lsim}{\raisebox{-1.5mm}{$\:\stackrel{\textstyle{<}}{\textstyle{\sim}}\:$}}

\def\r0{$\rho^{0}$}
\def\smb{{\small $\bullet$}}
\def\dcr{DCR$\phi $}
\newcommand{\rr}{r_{00}^{04}}

\newcommand{\sigmarho}{\hbox{$\sigma (\gamma p \rightarrow \rho^0 p)$}}
\newcommand{\sigmapi}{\hbox{$\sigma (\gamma p \rightarrow \pi^+\pi^- p)$}}
\newcommand{\mpipi}{M_{\pi \pi}}

\newcommand{\Wgp}{W_{\gamma p}}
\newcommand{\avWgp}{\langle\Wgp\rangle}
\newcommand{\avsgp}{\langle\sigma_{\gamma p} \rangle}
\newcommand{\avQ}{\langle Q^2\rangle}
\newcommand{\stat}{{\rm stat.}}
\newcommand{\syst}{{\rm syst.}}

\begin{document}

\begin{titlepage}
\begin{flushleft}
{\tt DESY 95-251    } \\
{\tt December 1995}\\

\end{flushleft}
\vspace*{4.cm}
\begin{center}
\begin{Large}
  \boldmath \bf{Elastic Photoproduction of $\rho^{0}$ Mesons at HERA\\}
  \unboldmath

  \vspace*{2.cm} H1 Collaboration \\ 
\end{Large}

\vspace*{1cm}

\end{center}

\vspace*{1cm}

\begin{abstract}
  The cross section for the elastic photoproduction of \r0\ mesons ($\gamma p
  \rightarrow \rho^0 p$) has been measured with the H1 detector at HERA for
  two average photon-proton centre-of-mass energies of 55 and 187~GeV\@.  The
  lower energy point was measured by observing directly the $\rho^{0}$ decay
  giving a cross section of $9.1\pm 0.9\,(\stat)\pm 2.5\,(\syst)\;\mu$b. The
  logarithmic slope parameter of the differential cross section, ${\rm
    d}\sigma/{\rm d}t$, is found to be $10.9 \pm 2.4\,(\stat) \pm
  1.1\,(\syst)\;$GeV$^{-2}$\@.  The \r0\ decay polar angular distribution is
  found to be consistent with s-channel helicity conservation. The higher
  energy cross section was determined from analysis of the lower part of the
  hadronic invariant mass spectrum of diffractive photoproduction and found
  to be $13.6\pm 0.8\,(\stat)\pm 2.4\,(\syst)\;\mu$b.

  \vspace*{1cm}

\end{abstract}
\end{titlepage}

\vfill
\clearpage
\begin{sloppypar}
\noindent
 S.~Aid$^{14}$,                   
 V.~Andreev$^{26}$,               
 B.~Andrieu$^{29}$,               
 R.-D.~Appuhn$^{12}$,             
 M.~Arpagaus$^{37}$,              
 A.~Babaev$^{25}$,                
 J.~B\"ahr$^{36}$,                
 J.~B\'an$^{18}$,                 
 Y.~Ban$^{28}$,                   
 P.~Baranov$^{26}$
% ...
\bigskip 

\noindent
{\footnotesize{
% ...
$ ^2$ III. Physikalisches Institut der RWTH, Aachen, Germany$^ a$ \\
% ...
$ ^{12}$ DESY, Hamburg, Germany$^a$ \\
% ...
$ ^{14}$ II. Institut f\"ur Experimentalphysik, Universit\"at Hamburg,
                                     Hamburg, Germany$^ a$  \\
% ...
$ ^{18}$ Institute of Experimental Physics, Slovak Academy of
                Sciences, Ko\v{s}ice, Slovak Republic$^ f$\\
% ...
$ ^{25}$ Institute for Theoretical and Experimental Physics,
                                                 Moscow, Russia \\
 $ ^{26}$ Lebedev Physical Institute, Moscow, Russia$^ f$ \\
% ...
$ ^{28}$ LAL, Universit\'{e} de Paris-Sud, IN2P3-CNRS,
                            Orsay, France\\
 $ ^{29}$ LPNHE, Ecole Polytechnique, IN2P3-CNRS,
                             Palaiseau, France \\
% ...
$ ^{36}$ DESY, Institut f\"ur Hochenergiephysik,
                              Zeuthen, Germany$^ a$\\
 $ ^{37}$ Institut f\"ur Teilchenphysik,
          ETH, Z\"urich, Switzerland$^ i$\\
% ...
$ ^a$ Supported by the Bundesministerium f\"ur
        Forschung und Technologie, FRG
        under contract numbers 6AC17P, 6AC47P, 6DO57I, 6HH17P, 6HH27I,
        6HD17I, 6HD27I, 6KI17P, 6MP17I, and 6WT87P \\
% ...
$ ^f$ Supported by the Deutsche Forschungsgemeinschaft\\
% ...
$ ^i$ Supported by the Swiss National Science Foundation\\
}}
\end{sloppypar}


\section{Introduction}

The photon, exhibiting both point-like and hadronic behaviour, has a rich
structure that is still not yet fully explored.  New opportunities for
studying photon interactions are opened by the HERA electron-proton collider.
The nearly real photons produced by HERA's electron beam extend the energy
range for photoproduction processes by more than one order of magnitude, up
to the equivalent of tens of TeV in fixed target experiments, providing a
large kinematic range for studying the properties of the photon. The H1 and
ZEUS collaborations have begun to explore this domain, e.g.\ with
measurements of the total $\gamma p$ cross section~\cite{H1gp,ZEUS1,NewH1}
and vector meson production~\cite{ZEUS1,JPSIH1,JPSIZEUS,ZEUS2,PHIZEUS}.

In this paper we report the H1 results for the elastic \r0\ photoproduction
cross section at two average total energies, $\avWgp=55$ and 187~GeV, of the
$\gamma p$ centre-of-mass system (cms).  For the low energy measurement the
associated differential cross sections, ${\rm d}\sigma/{\rm d}\mpipi$, where
$\mpipi$ is the invariant mass of the $\rho^0$ decay products, ${\rm
  d}\sigma/{\rm d}t$, where $t$ is the square of the four-momentum transfer
at the proton vertex, and the polar decay angle distribution, ${\rm
  d}\sigma/{\rm d}\cos \theta ^*$, of the \r0\ are also presented.

At lower energies the photon is known to exhibit behaviour resembling
that of a hadron. In particular, the reaction
\begin{equation}
  \gamma + p \rightarrow \rho^0 + p
  \label{eqn:elrho}
\end{equation}
at cms energies of 10-20~GeV behaves much like elastic scattering of hadrons:
the cross section is roughly constant with cms energy, the dependence on $t$
of ${\rm d} \sigma/{\rm d}t$ is close to exponential and $s$-channel helicity
conservation holds (the \r0 retains the helicity of the photon).  This is
interpreted in the Vector Meson Dominance (VMD) model~\cite{VMD1,BSYP78} as
the fluctuation of the photon into a virtual \r0\ meson, with subsequent
elastic scattering of the \r0\ on the proton.  Regge theory~\cite{Collins}
inspired models have been used successfully to describe the energy variation
of the total cross section, with a `soft' pomeron term dominating at high
energies~\cite{SchSjoNUCB}\@.  Models based on perturbative quantum
chromodynamics (pQCD) may combine a soft VMD component with perturbatively
calculable components to arrive at a total $\gamma p$ cross section.  The
$\gamma p \rightarrow \rho^0 p$ cross section may then be completely
determined by the parameters used to fit the total $\gamma p$ cross section.
Thus studies of reaction (\ref{eqn:elrho}) at HERA energies help to constrain
such models, as well as to clarify the nature of the pomeron. Thorough
exploration of these interactions in the HERA energy regime is important for
an eventual understanding of the hadronic properties of the photon.

Almost real photons, radiated at small angles by HERA's electron beam
(26.7~GeV electrons in 1993 and 27.5~GeV positrons\footnote{The term
  `electron' will subsequently be used to refer to both electrons and
  positrons.} in 1994), can interact with the 820 GeV protons at very high
energies in the $\gamma p$ cms. In the VMD model, one expects dominant
production of the vector meson series $\rho^{0}$, $\omega$, $\phi$ along with
other hadronic production.  Elastic $\rho^{0}p$ interactions are
characterised by a scattering of the virtual $\rho^{0}$ meson and the proton
at low $|t|$, leading to a $\rho ^0$ at low $p_{t}$ relative to the HERA beam
axis.  The $\rho^{0}$ subsequently decays into a $\pi ^{+} \pi ^{-}$ pair.

The kinematic variables are defined as follows:
\begin{eqnarray*}
  Q^2 & = & -(k-k^{\prime})^2   =  -q^2 \\
   t  & = & (q-v)^2  =  (P-P^{\prime})^2 \\
   y  & = & {q\cdot P \over k \cdot P} \\
   s  & = & (P + k)^2 \\
   W_{\gamma p} & = & \sqrt{y s}
\end{eqnarray*}
where $y$ is the inelasticity parameter, $Q^2$ is the negative square of the
four-momentum transfer at the electron vertex, $k (k^{\prime})$ denotes the
four vector of the incoming (scattered) electron, $P (P^\prime)$ of the
incoming (scattered) proton, $q$ of the photon and $v$ of the vector meson.

The two measurements presented in this paper were made under very different
experimental conditions.  In the low $\Wgp$ analysis, where the \r0\ rest
frame is boosted by about 1 unit of rapidity toward the electron beam
direction, the \r0\ decay products were observed in the central tracking
detector (see section~\ref{sec:h1det} for details of the H1 detector),
enabling good reconstruction of the hadronic final state (excluding the
proton, unseen in both analyses).  The $\Wgp$ range ($40 < \Wgp< 80$~GeV) is,
however, too low for the scattered electron to be detected in the small angle
electron tagger.  The kinematic quantities were therefore reconstructed from
the {\r0}'s two decay pions, using the assumptions of an elastic interaction
and of $Q^2 = 0$ GeV$^2$\@.  The resonant contribution to the cross section
is deduced from the analysis of the line shape.

The high $\Wgp$ analysis used those data with an electron tag so that the
$\Wgp$ range was $164 < \Wgp < 212$~GeV\@.  At these high values of $\Wgp$
the \r0\ rest frame is boosted by about 3 units of rapidity toward the
electron beam direction and the decay products are kinematically forced
outside the acceptance region of the tracking detectors, requiring
reconstruction of the hadronic final state in the calorimeters.  Since the
spatial and energy resolutions of the calorimeters are worse than those of
the trackers no measurement was possible of the resonance line shape, $t$
dependence, or polar decay angle distribution.

The measured $ep$ cross section is converted into a $\gamma p$ cross section,
averaged over the available $\Wgp$ and $Q^2$ range so that $\avsgp = \int
\frac{{\rm d^2}\sigma_{ep}}{{\rm d}y{\rm d}Q^2}\,{\rm d}y{\rm d}Q^2 / \int
\Phi(y,Q^2) \,{\rm d}y {\rm d}Q^2$.  In order to evaluate the photon flux,
$\Phi(y,Q^2)$, we ignore any lepton beam polarisation and the longitudinal
(helicity zero) contribution is neglected. 
In this case the flux for transverse virtual photons is given by the
Weizs\"acker-Williams approximation which we take in the form:
\begin{eqnarray} 
  \label{eq:ijray}
  \Phi\left(y,Q^2\right) = {\alpha\over2\pi Q^2}\left({1+(1-y)^2\over
    y} - {2(1-y)\over y}\cdot{Q^2_{\rm min} \over Q^2}\right)
  \;\;;\;\;
  Q^2_{\rm min} = {m_e^2 y^2 \over 1-y},
\label{eqn:wwa}
\end{eqnarray}
where $\alpha$ is the fine structure constant and $m_e$ is the electron mass
~\cite{BUROW,WW}\@.  The quantity $Q^2_{\rm min}$ is the minimum photon
virtuality which is kinematically allowed.  The upper limit on the $Q^2$
range over which the measurements were taken, $Q^2_{\rm max}$, is determined
by the experimental procedure used.  For the low $\Wgp$ analysis the
requirement of $|t_{\rm rec}|<0.5$~GeV$^2$ (see section~\ref{sec:lowWgpana})
implies effectively $Q^2_{\rm max} \approx 0.5$~${\rm GeV}^2$ and the mean
$Q^2$ is $\avQ = 0.035$~${\rm GeV}^2$\@.  The requirement of the scattered
electron to be detected in the tagger in the high $\Wgp$ analysis gives
$Q^2_{\rm max} \approx 0.01$~${\rm GeV}^2$ and $\avQ = 0.001$~${\rm GeV}^2$.

With an assumption for the $Q^2$ dependence of the virtual photon-proton
cross section, the total elastic $\rho^0$ cross section at $Q^2=0$ may be
extrapolated from $\avsgp$\@.  In a VMD model~\cite{VMD1} in which the
$\rho^0$ dominates, the longitudinal and transverse virtual photon-proton
cross sections are related by $\sigma^L_{\gamma^{*} p}=\sigma^T_{\gamma^{*}
  p} \cdot Q^2/M^2_\rho$ and evolve such that $\sigma^T_{\gamma^{*} p}
\propto 1/ (1+Q^2/M^2_\rho)^2$\@.  With these assumptions, corrections
averaged over the selected samples of events amount to $+3.9\;\%$ and
$+0.2\;\%$ for the low and high $\Wgp$ analyses respectively.  All results
quoted in this paper do not include any correction for this extrapolation.


\section{The H1 Detector}
\label{sec:h1det}

A complete description of the detector is available elsewhere~\cite{H1det}\@.
Only a brief description of the parts of the detector relevant to these
analyses is included here. In H1 the $z$-axis points in the proton beam
direction, denoted by the term `forward'.

The H1 Central Tracking Detector (CTD) provides charged track measurement and
triggering in the pseudorapidity ($\eta = -\ln\tan\theta/2$) range
$-1.5<\eta<1.5$\@.  Its main components are two concentric drift chambers,
Central Jet Chambers 1 and 2 (CJC1 and CJC2)\@.  Located at the inner radius
of CJC1 and CJC2 are two sets of cylindrical drift chambers for measurement
of the $z$-coordinate and two proportional chambers, CIP and COP\@.  The
entire CTD is situated within a 1.15~T solenoidal magnetic field, parallel to
the proton beam axis.  The transverse momentum resolution for charged tracks
is $\sigma(p_t)/p_t \approx 0.009 \cdot p_t ({\rm GeV}) \oplus 0.015$\@. The
CTD was the detector used to trigger and reconstruct the two particle
hadronic final state in the low $\Wgp$ analysis.

The CTD is surrounded by a finely segmented Liquid Argon calorimeter (LAr)
that provides hadronic and electromagnetic energy measurements in the range
$-1.5<\eta<3.3$\@.  Instrumented sections of the return yoke of the magnet
outside the LAr calorimeter provide tracking and hadronic `tail-catching'
calorimetry in the range $-2.5<\eta<3.4$\@. These detectors were used to veto
inelastic and cosmic muon events.

In the forward region two detectors were used to veto proton dissociation
events: the forward muon detector (FMD) and the proton remnant tagger
(PRT)\@.  They were used in the high $\Wgp$ analysis to tag hadrons produced
at large values of pseudorapidity ($5.0<\eta<7.5$) by detecting secondary
particles issued from collisions with the beam pipe or adjacent
material.  The FMD consists of planes of drift chambers at $6<z<9$~m
and the PRT is made up of scintillator planes at $z=24$~m.

Two detectors provide coverage in the backward direction.  The Backward
Electromagnetic Calorimeter (BEMC), a lead scintillator calorimeter
covering the range $-3.4<\eta<-1.5$, was used to reconstruct the hadronic
final state in the high $\Wgp$ analysis, while the Backward Proportional
Chamber (BPC), $-3.0<\eta<-1.5$, helped in the rejection of beam-gas
background.

Behind the BEMC are located the Time of Flight (ToF) scintillator walls,
covering a pseudorapidity range of $-3.5<\eta<-2.0$\@.  Their timing
information was used to reject background upstream proton interactions and in
the high $\Wgp$ analysis as a trigger.

The luminosity system, monitoring the reaction $ep \to e\gamma p$,
consists of two TlCl/TlBr crystal calorimeters.  The electron tagger is
located at $z=-33$~m and the photon detector at $z=-103$~m.  The electron
tagger was also used to detect and trigger on the scattered electron for
the high $\Wgp$ analysis.


\section{Monte Carlo}

Two Monte Carlo (MC) generators were used in these analyses,
PYTHIA~\cite{PYTHIA} and PHOJET~\cite{phojet}\@.  Five MC photoproduction
mechanisms are distinguished: elastic interactions (EL) which describe
$\gamma p \rightarrow Vp$ where $V$ stands for one of the vector mesons
$\rho^0$, $\omega$ or $\phi$, single photon diffractive dissociation (GD) in
which the photon dissociates ($\gamma p \rightarrow Xp$), single proton
diffractive dissociation (PD) in which the proton dissociates ($\gamma p
\rightarrow VY$), double diffractive dissociation (DD) in which both photon
and proton dissociate ($\gamma p \rightarrow XY$) and non-diffractive
interactions (ND)\@.  This last category includes all reactions not included
in the previous classes.  The ratios of the cross sections for the 5
processes were chosen following comparisons made using the high $\Wgp$ data
(see section~\ref{highwanal}) and the constraint imposed by the measurement
made of the total $\Wgp$ cross section~\cite{NewH1}:

\begin{equation}
  \sigma_{\rm EL}:\sigma_{\rm GD}:\sigma_{\rm PD}:\sigma_{\rm DD}:\sigma_{\rm ND}
  = 1 : 1.25 : 0.75 : 2.00 : 5.0 \, .  \label{eq:ratios}
\end{equation}

All channels were generated with a $Q^2$ dependence following 
equation~\ref{eq:ijray} with no dependence on $\Wgp$ for the
photon-proton cross section.

Both MC generators have similar inputs for the EL channel.  Events are
generated with a $t$-dependence of $e^{bt}$ where $b \approx 11$~${\rm
  GeV^{-2}}$, a decay angular distribution in the $\rho^0$ rest frame $\cos
\theta ^*$ proportional to $\sin^2 \theta^*$, as expected for a transversely
polarised $\rho^0$ (helicity $\pm 1$), and a skewed mass distribution
consistent with that measured (see section~\ref{results})\@.  The relative
rates of $\rho^0$, $\omega$ and $\phi$ production were chosen to be 13:1.5:1
following measurements made at lower energies~\cite{BSYP78}.

For the singly diffractive dissociation channels (GD, PD) PYTHIA assumes a
$t$-dependence of $b \approx 5$~${\rm GeV^{-2}}$\@.  PHOJET assumes a $b$
parameter that depends on the mass of the diffractive system, $M_X$, so that
$b \approx 11$~${\rm GeV^{-2}}$ at $M_X=M_\rho$ and becomes smaller for
higher $M_X$~\cite{engel}\@.  For the double diffractive dissociation and for
both MC generators, $b \approx 2-3\;{\rm GeV^{-2}}$.

Diffractive dissociation of the proton or photon is described in PYTHIA
according to a distribution such that roughly ${\rm d} N/{\rm d}M_X^2 \sim
1/M_X^2$ with an enhancement at lower masses.  Masses are generated starting
at 0.2~GeV (twice the pion mass) above the mass, $M_{in}$, of the incoming
particle, with the $\rho^0$ mass used for the incoming $\gamma$\@.  PHOJET in
contrast generates a mass spectrum according to ${\rm d}N/{\rm d}M_X^2 \sim
1/(M_X^2-M_{in}^2)$ at higher masses and treats the lower part as a single
resonance.  The spectrum starts at twice the pion mass above $M_{in}$\@.

The MC event samples included a full simulation of the H1 detector and
were subject to the same reconstruction as each of the two data samples.


\section{The Low {\boldmath $\Wgp$} Analysis}
\label{sec:lowWgpana}

The data for this analysis were taken in 1993 when HERA operated with 90
bunches of 820~GeV protons and 94 bunches of 26.7~GeV electrons with 84
colliding bunch pairs.  A small number of non-colliding electron and proton
bunches, called `pilot' bunches, were included in order to estimate the
beam-gas background. The data sample was limited to the largest continuous
segment of runs containing stable and uniform central tracker conditions and
amounted to a total integrated luminosity of $19.8 \pm 1.0$~$\rm{nb}^{-1}$\@.

The most favourable trigger capable of distinguishing the soft, low
multiplicity hadronic final state from the high level of background was the
Drift Chamber $r$-$\phi$ trigger (\dcr) which distinguished individual tracks
by matching CJC1 and CJC2 wire hits with predefined masks~\cite{dcrphi}\@.
The major drawback to this trigger was that its efficiency threshold, at
$p_t\approx 450$~MeV, was located above the peak of the decay pion $p_t$
distribution.  It was found that the best trigger condition required one and
only one track pattern identification by the {\dcr}\@.  A requirement of two
(or more) tracks above the threshold would have severely reduced the trigger
efficiency and resulted in an unacceptable rate of background triggers.
Thus, the second track in the event would usually have a $p_t$ below
threshold and would not be detected by the trigger.  In order to reduce the
rate of beam-gas triggers, the overall trigger required in addition a single
vertex reconstruction by CIP and COP~\cite{zvtx} ($z$-vertex) and was vetoed
if there was significant energy present in the forward part of LAr, excessive
activity in the backward portion of CIP or an out-of-time ToF signal
indicating a background event.

Approximately $1.4\times 10^5$ events satisfied the trigger. Analysis of the
data from the pilot bunches indicated that the vast majority of  triggered
events  consisted of proton beam induced background. The data were reduced in
a three stage procedure consisting of computer selections and a visual scan.

Since this analysis relies on the identification of the two particle
final state, it is particularly sensitive to the quality of track
reconstruction.  Having observed differences between the data and MC track
multiplicities, a visual scan early in the analysis was found to be necessary
in order to establish equivalent data and MC samples of clean two particle
events. This procedure also allowed a measurement of the event reconstruction
efficiency independent of the MC simulation.

The first stage in data reduction employed a loose track requirement which
was essentially independent of track reconstruction quality. Its purpose was
to eliminate the obvious events that were not composed of only two particles,
thereby reducing the data sample to a manageable size for scanning. Those
events with a total of 1 track and no other track segments, 2 like sign, or
more than 3 well reconstructed tracks were rejected.  All other events were
kept. A sub-sample of 600 of the rejected events were scanned and only one
event was  incorrectly rejected due to an improper track reconstruction
(a $0.3\, \%$ error).

The reduced sample of $\approx 47 \times 10^3$ events was then scanned using
the H1 event display program. Each event was required to have two opposite
sign track patterns pointing to a common vertex and extending radially out to
at least CJC2 ($r \stackrel{>}{_{\sim}}43$~cm)\@.  Only digitised hit
information from the CJC detectors was used in the scan.  Obvious cosmic
events with tracks in the muon detector were removed. A total of 12,200 valid
two-particle, opposite-sign events passed the scan including 623 from the
proton pilot bunch crossings and 13 from the electron pilot bunch crossings.

Using the track fitting of the standard H1 reconstruction program, the third
stage of the reduction focussed on filtering two-particle events with
well-reconstructed vertices. The efficiencies for data and MC events to pass
this stage in the analysis were found to be $0.8430\pm 0.0034\,(\stat)$ and
$0.8807\pm 0.0022\,(\stat)$, respectively.  The ratio of the data to MC
efficiencies was used as a correction factor to account for this remaining
difference in event reconstruction ($C$ in equation~\ref{eqn:dsdm})\@.

\begin{figure}
  \begin{center}
    \epsfig{file=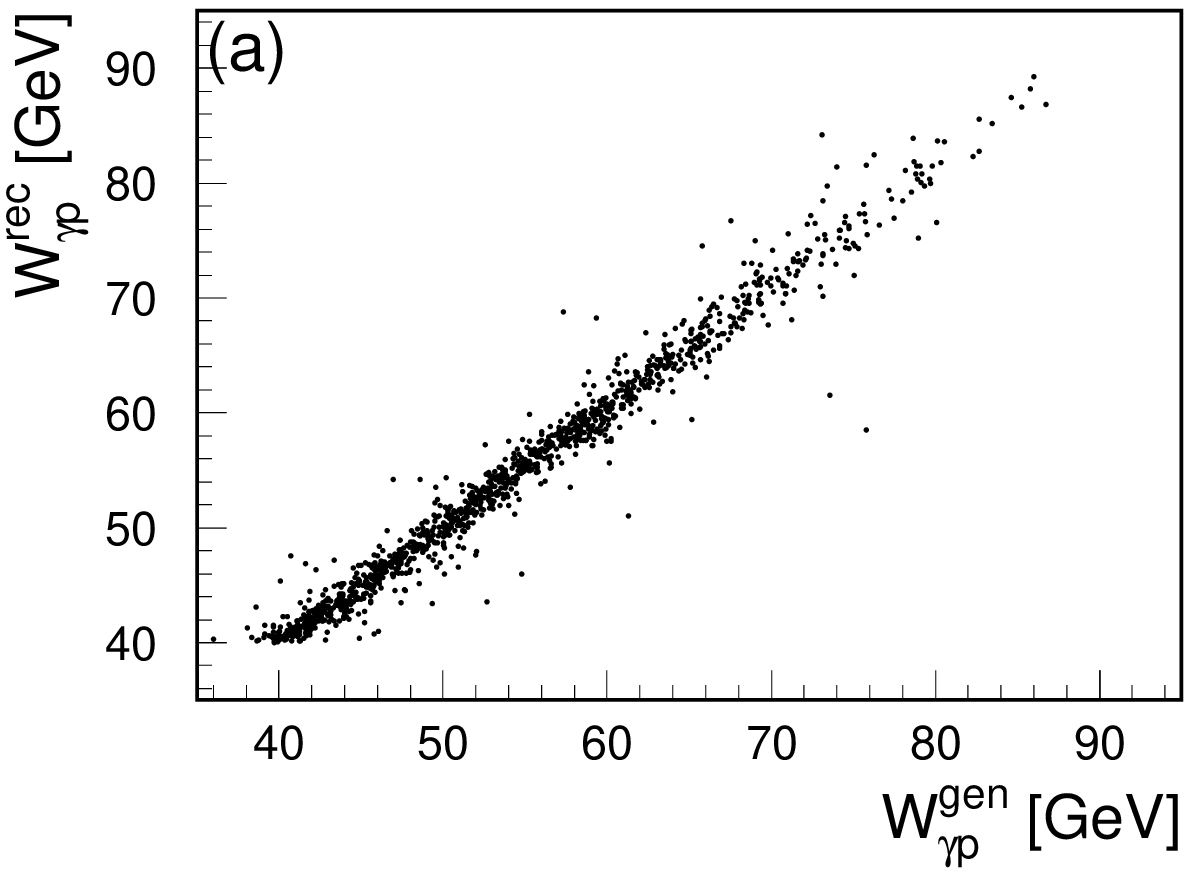,width=0.49\textwidth}
    \epsfig{file=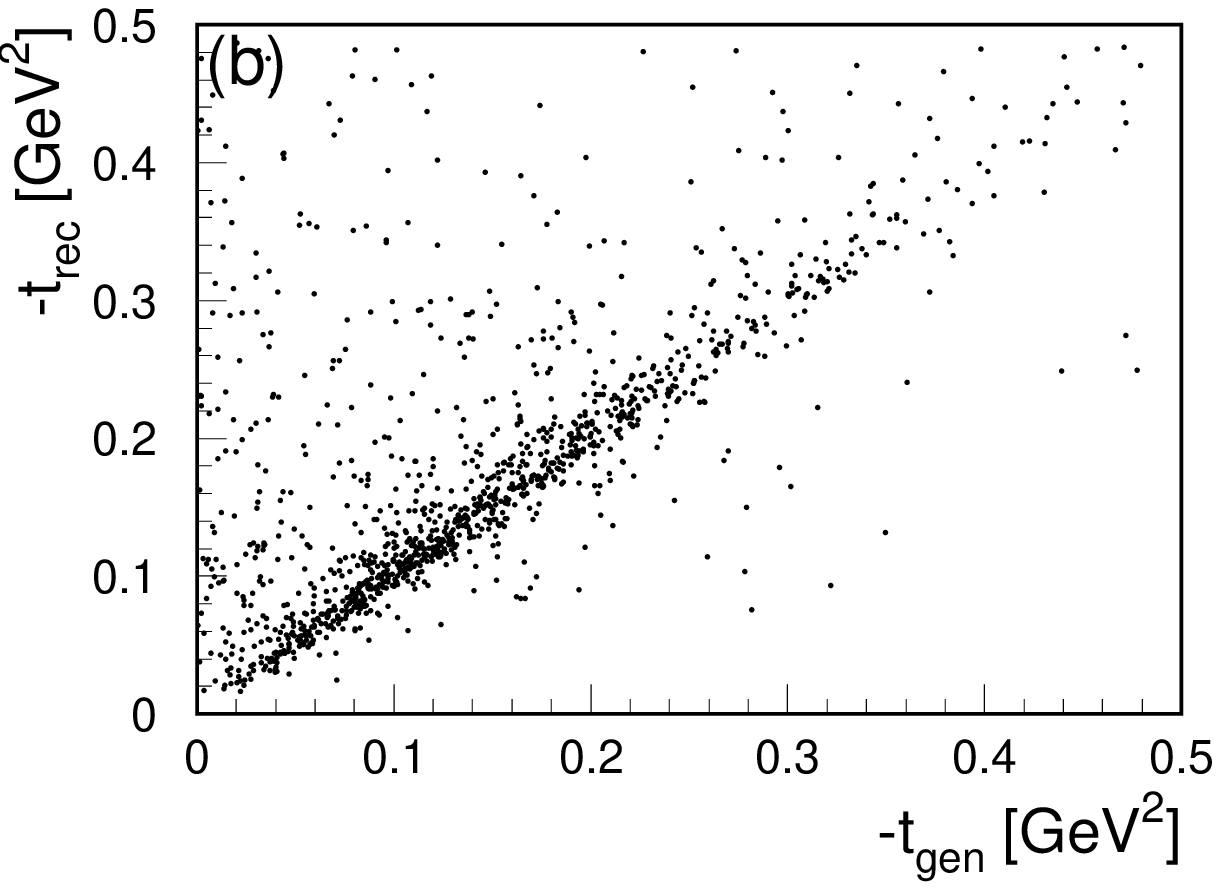,width=0.49\textwidth}
  \end{center}
  \caption {\small 
    {\rm (a)} MC reconstructed versus generated $W_{\gamma p}$; {\rm (b)} MC
    reconstructed versus generated $t$ for the low $\Wgp$ analysis. Only
    elastic $\rho^0$ events are shown.  The distributions are shown after all
    cuts were applied.}
  \label{fig:Wgpstudy}
  \label{fig:tstudy}
\end{figure}

An accurate reconstruction of kinematic variables was possible using only the
reconstructed \r0\ and the assumption of an elastic event with $Q^2 =
0$~GeV$^2$\@.  Figures~\ref{fig:Wgpstudy}a and \ref{fig:tstudy}b compare the
reconstructed to the generated values for $W_{\gamma p}$ and $t$\@.  The mean
differences were not significant: $\Delta \Wgp \approx 0.4$ GeV with an RMS
of 1.2 GeV, $\Delta t \approx 0.0$ GeV$^2$ with an RMS of $0.02$ GeV$^2$\@.

The final set of filters uses kinematic cuts to reduce the number of
background events. Each event was required to have:
\begin{enumerate}
\item[\smb] a total energy in the forward, inner part of the LAr calorimeter
  ($1.7<\eta<3.3$), of less than 400~MeV suppressing events in which the
  proton dissociates,
\item[\smb] $p_t>0.5$~GeV for the high $p_t$ track and $p_t<0.4$~GeV for the
  low $p_t$ track, matching the region of flat efficiency in $p_t$ for the
  \dcr\ trigger,
\item[\smb] $|t_{\rm rec}|<0.5\;{\rm GeV^2}$, with $t_{\rm rec} =
  -p_t^{\pi\pi}$ being the reconstructed $t$, reducing the number of
  dissociative and inelastic events,
\item[\smb] $\Wgp^{\rm rec}>40\;{\rm GeV}$, cutting most of the proton
  beam-gas events present in data and restraining the measurement to the
  range where there is reasonable acceptance,
\item[\smb] $M_{\pi\pi}\in[0.52,1.17]$~GeV, limiting the measurement to the
  range where there is a reasonable signal to background ratio.
\end{enumerate}
\noindent A total of 358 colliding bunch events and no pilot events survive
all cuts.

The trigger efficiency was measured using independent samples of events,
where there was a chance overlap between the reaction $ep \to e\rho^0 p$ and
the Bethe-Heitler process $ep \to e\gamma p$\@.  These events were triggered
by the detection of the electron from the Bethe-Heitler process in accidental
coincidence with a signal from other detector components.  The LAr veto was
found to be 100\% efficient for the selected events.  Triggers independent of
the CTD were used to determine the $z$-vertex efficiency $\epsilon_{z-\rm
  vertex}$\@.  With slightly relaxed selection criteria, $\epsilon_{z-{\rm
    vertex}} = 0.72 \pm 0.03$(\stat) was obtained, in agreement with MC
simulations.  The relative efficiency of the \dcr\ and remaining requirements
for events already fulfilling the $z$-vertex trigger, $\epsilon_{\mbox{\rm
    \tiny \dcr}}$, was determined from a sample triggered by the electron and
the $z$-vertex condition only.  Applying the final selection procedure and
cuts ensured the same event properties as for the signal sample and gave
$\epsilon_{\mbox{\rm \tiny \dcr}} = 0.29 \pm 0.06$(\stat)\@.  The results
were multiplied together to obtain the overall trigger efficiency of
$\epsilon_{\rm trig} = 0.21\pm0.04$\@.  The error quoted is dominated by the
low statistics in the overlap sample.

Backgrounds in the final sample were estimated using the PYTHIA MC generator.
Only the diffractive channels were found to contribute significantly.  The
estimated background fractions in the final sample of $\rho^0$ candidates was
$13\pm 7 \, \%$ from GD, $9\pm 5 \, \%$ from PD, $10\pm 5 \, \%$ from DD, $0
\pm 1 \, \% $ from ND and $1 \pm 1 \, \%$ from elastic production of $\omega$
and $\phi$ mesons\footnote{The $p_t$ cuts imposed on the tracks strongly
  suppress events from the reaction $\gamma p \rightarrow \phi p$, $\phi
  \rightarrow K^+ K^-$}\@.  The errors on the background includes the
uncertainty in the decomposition of the total photoproduction cross section.

\begin{figure}[hbt]
  \noindent \epsfysize=4truein \centerline{\epsfbox{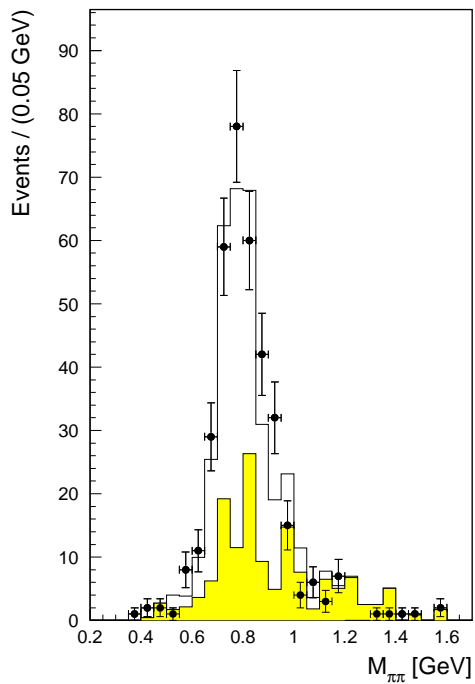}} \caption
  {\small Reconstructed $M_{\pi \pi }$ distributions for the low $\Wgp$ data
    sample before background subtraction and acceptance correction.  The open
    histogram shows the MC (PYTHIA) prediction for all possible channels
    (normalised to the data).  The shaded histogram shows the MC prediction
    for all channels apart from EL.}
  \label{fig:final}
\end{figure}

Figure~\ref{fig:final} shows the dipion mass spectra of the accepted data
(without background subtraction) for the low $\Wgp$ analysis along with the
MC prediction.  The MC mass shape, containing a Ross-Stodolsky skewed mass
distribution for the elastic \r0\ events (see section~\ref{massspectrum}
for details), matches the shape of the data histogram well.

\section{The High {\boldmath $\Wgp$} Analysis}
\label{highwanal}

The data for this measurement were collected during a short dedicated run in
the 1994 data taking period with a vertex shifted forward of its nominal
position by $\approx70$~cm.  With this vertex position the acceptance for
particles travelling backwards at small angles to the beam pipe is enhanced.
The HERA machine was operated with 153 colliding bunches of 27.5~GeV
positrons and 820~GeV protons with in addition 17 proton and 15 positron
pilot bunches.  The data taken correspond to a total integrated luminosity of
$23.8 \pm 1.3$~${\rm nb}^{-1}$\@.

Photoproduction events were triggered by the coincidence of a signal in the
electron tagger ($E_e^\prime > 4$~GeV) and a signal in the ToF system coming
from the time interval expected for $ep$ interactions. This trigger has been
shown to have a high ($35\,\%$) and well understood acceptance for all
photoproduction processes for data taken over a similar $\Wgp$
range~\cite{NewH1}\@.

The main source of background in the triggered sample was found to be
electron beam interactions with residual gas or with material inside the
beam pipe.  This background was reduced by requiring at least one BPC hit and
a total energy deposition greater than 0.2~GeV in the BEMC or LAr
calorimeters.  The measurement was further restricted to the kinematic range
$0.3<y<0.5$ where there was good electron tagger acceptance.

Two classes of diffractive events were distinguished on the basis of a
`rapidity gap' i.e.\ a region of laboratory pseudorapidity where no particles
are observed in the final state.  The first, termed an elastic proton sample
where in most cases the proton remained intact (GD and EL), was selected as
follows. No activity was required above threshold in the forward detectors
(FMD and PRT)\@.  The pseudorapidity, $\eta_{\rm max}$, of the most forward
energy deposit greater than 600~MeV in the LAr or BEMC calorimeters or the
most forward track in the CTD was required to be less than 3\@.  The second
class of event, where the proton dissociates (DD and PD), was selected by
requiring there to be signals in either FMD, PRT or $\eta_{\rm max}>3$\@.  ND
events in the proton dissociation sample were suppressed by requiring $\Delta
\eta$, the largest region in the detector where no tracks or energy deposits
were found, to be greater than 2 units of pseudorapidity with the upper edge
of the rapidity gap having $\eta>2.5$\@.  It should be noted that these
latter requirements also suppress those events where the proton fragments
into a high mass system and in which the rapidity gap is small or
non-existent. The distinction between these DD events and ND events is
experimentally not well defined.  Studies with the MC generator indicate that
these cuts restrict the proton dissociation sample to a proton dissociation
mass of $M_{Y}<10$~GeV\@.

\begin{figure}[hbt]
 \begin{center}
   \epsfig{file=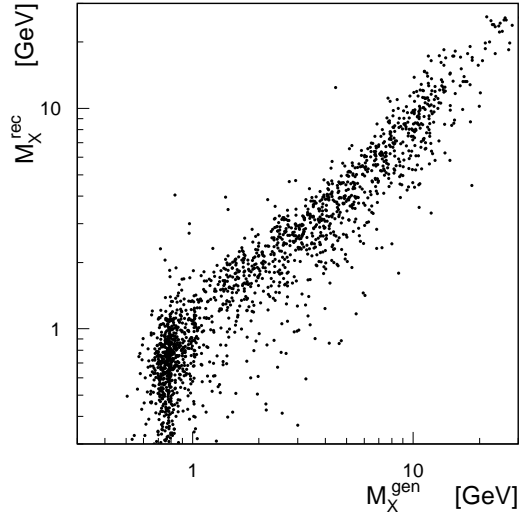,width=0.5\textwidth}
   \caption{\small MC reconstructed versus generated  $M_X$ 
     for the high $\Wgp$ analysis. All diffractive channels (EL, GD, DD, GD)
     are shown. }
   \label{mxgenrec}
 \end{center}
\end{figure}

The four vector $(E_h, P_{xh}, P_{yh}, P_{zh})$ of the hadronic final state
excluding the proton or the dissociative proton system was determined by
combining calorimeter (LAr and BEMC) and tracking (CTD) information in a
procedure that avoided double counting and optimised the resolution.  For the
proton dissociation sample, only those tracks/energy deposits at angles more
backward of the rapidity gap were included.  The reconstructed invariant mass
of the hadronic final state excluding the proton, $M^{\rm rec}_X$, was then
calculated making the assumption that $P^2_{xh}+P^2_{yh} \simeq |t| \ll
(M^{rec }_X)^2$ and using the approximation $y=(E_h-P_{zh})/(2E_e)$ so that
\begin{eqnarray}
  M^{\rm rec}_X=\sqrt{(E_h+P_{zh})\cdot 2\cdot E_e \cdot y}.
\end{eqnarray}
Here $E_e$ is the electron beam momentum and $y$ is reconstructed from the
final state electron.  This method resulted in good resolution, as
demonstrated by figure~\ref{mxgenrec}, even though a significant proportion
of the hadronic energy is not detected in the backward region.  It should be
noted, however, that the resolution width in $M^{\rm rec}_X$ is much greater than
the width of the $\rho^0$ decay, making it impossible to determine the line
shape with this analysis.

A sample of events where elastic $\rho^0$ production dominates was selected
by requiring $0.4<M_X<1.26$~GeV in the elastic proton sample. 

Background in the elastic $\rho^0$ sample was observed and attributed to four
dominant sources: electron beam induced, GD, PD where none of the proton
fragments were detected and EL from $\omega$ and $\phi$ production.

The electron beam induced background was estimated using pilot bunch events,
was found to contribute $\approx 20 \pm 5\, \%$ and was removed from the
sample.

\begin{figure}[hbt]
  \begin{center}
    \epsfig{file=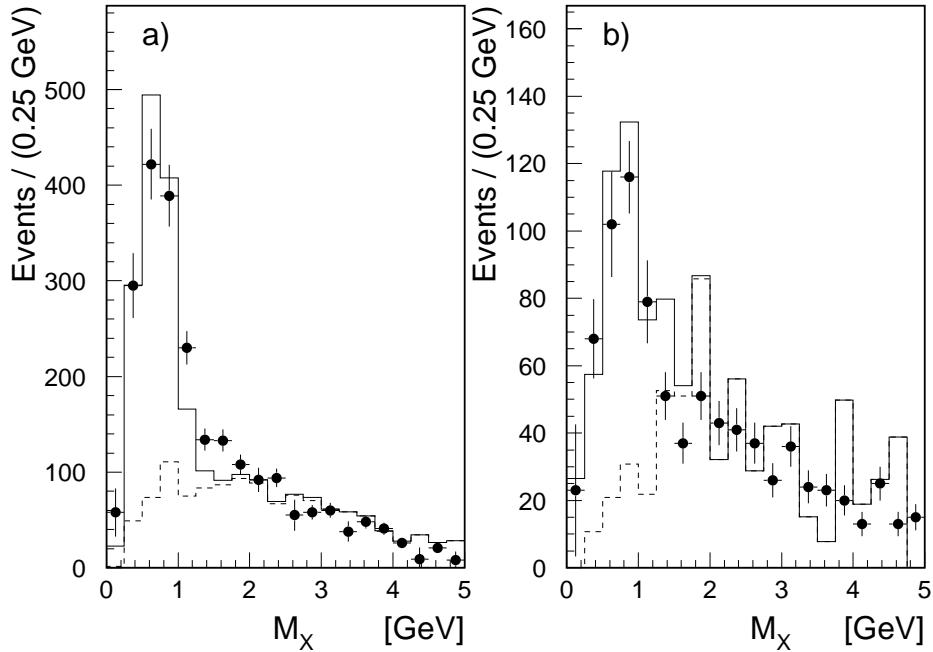,width=0.85\textwidth}
    \caption{\small  Reconstructed $M_X$ distribution for the
      high $\Wgp$ data for a) the elastic proton selection and b) for the
      proton dissociation selection. The solid histograms show the MC
      prediction for all possible channels. The dashed histograms show the MC
      prediction for a) all channels apart from EL and b) all channels apart
      from PD\@.  The MC distributions were normalised to the data in the
      region $M_X<1.26$~GeV in the elastic proton selection. }
    \label{rhomxei}
  \end{center}
\end{figure}

The MC generators (PYTHIA and PHOJET) were used to estimate the fraction of
GD, PD, EL, DD and ND background. The relative cross sections used in the
generation for the four diffractive processes described above were chosen by
comparison of the $M^{\rm rec}_X$ spectra in data and MC for both the elastic
proton and proton dissociation samples.  The region where $M^{\rm
  rec}_X>1.26$~GeV was found to be dominated by GD and DD and that where
$M^{\rm rec}_X<1.26$~GeV by EL and PD\@.  Since the relative cross sections
obtained in this comparison for PYTHIA and PHOJET were in broad agreement,
the same fractions were used for both MC generators (see
equation~\ref{eq:ratios})\@.  This choice also satisfied the constrains
imposed in~\cite{NewH1}\@.  Figures~\ref{rhomxei} (a,b) show comparisons of
the data with MC predictions (with the cross section ratios as described
above) for the $M^{\rm rec}_X$ distributions of the elastic proton and
dissociated proton samples. For these plots and subsequent results the
average of the two MC generators was used.

The ratio of GD events produced in the MC generators for $0.4<M^{\rm
  true}_X<1.26$~GeV to EL events was found to be $\approx 6\,\%$\@.  This
number agrees well with measurements made at lower
energies~\cite{goulinanos}\@.  The GD, PD, EL, DD and ND background fractions
in the elastic $\rho^0$ sample were estimated to be $12\pm 6\,\%$, $6\pm 3\,
\%$, $10\pm 5\,\%$, $3\pm 2\,\%$ and $0\pm 1\,\%$ respectively.  As for the
low $\Wgp$ case the errors on the background estimates accommodate the range
of uncertainty in the decomposition of the total photoproduction cross
section into its five constituent processes~\cite{NewH1}\@.  All backgrounds
were subtracted statistically from the data.

The acceptance for the electron tagger and for the detection of the hadronic
final state were determined separately.  The first was determined in a
procedure identical to that described in~\cite{NewH1}, was found to be $61\,
\%$\@.  The hadronic final state acceptance was determined by taking the mean
of the values obtained from two MC generators, namely $39\, \%$ and $40\, \%$
for PYTHIA and PHOJET respectively.


\section{Results}
\label{results}

\subsection{The  Differential {\boldmath $\pi^+ \pi^-$} Cross Section
  {\boldmath ${\rm d}\sigma/{\rm d}\mpipi$}}
\label{massspectrum}

The corrected differential cross section of the invariant mass of the
$\pi^+\pi^-$ system, ${\rm d}\sigma/{\rm d}\mpipi$, was determined for the
data taken at $\avWgp=55$~GeV from the raw distribution (shown in
figure~\ref{fig:final}) using
\begin{eqnarray}
  \frac{{\rm d}\sigma}{{\rm d}\mpipi}=\frac{N_{\rm bin}-N_{\rm bgd}}{\Phi_I \, {\cal L}
    \, \varepsilon_{\rm trig} \, \varepsilon_{\rm bin} \, C} \cdot
  \frac{1}{1+R_{\rm PD}} \cdot \frac{1}{\Delta M}.
\label{eqn:dsdm}
\end{eqnarray}
Here:
\begin{list}{-}{
\itemsep 0cm
\parsep  0cm
\topsep  0.2cm
}
\item $N_{\rm bin}$ is the number of events reconstructed in each mass bin,
\item $N_{\rm bgd}$ is the estimated number of background events in the bin,
  from GD, DD, ND and EL,
\item $R_{\rm PD}$ is the estimated ratio of PD to EL (taken as a constant
  scaling factor for all bins),
\item $\Phi_I$ is the photon flux found by integration of
  equation~\ref{eq:ijray}, for $40<\Wgp<80$~GeV and $Q^2_{\rm
    min}<Q^2\lsim0.5\;{\rm GeV^2}$,
\item ${\cal L}$ is the integrated luminosity,
\item $\varepsilon_{\rm trig}$ is the global trigger efficiency as calculated
  in section~\ref{sec:lowWgpana},
\item $\varepsilon_{\rm bin}$ is the bin by bin selection efficiency given
 by the
  MC before the trigger selection, averaged over the kinematical range
  generated, 
\item $C$ is the efficiency correction equal to the ratio of data to MC track
  reconstruction efficiencies,
\item $\Delta M$ is the bin width.
\end{list}

\begin{figure}[hbt]
 \begin{center} \epsfig{file=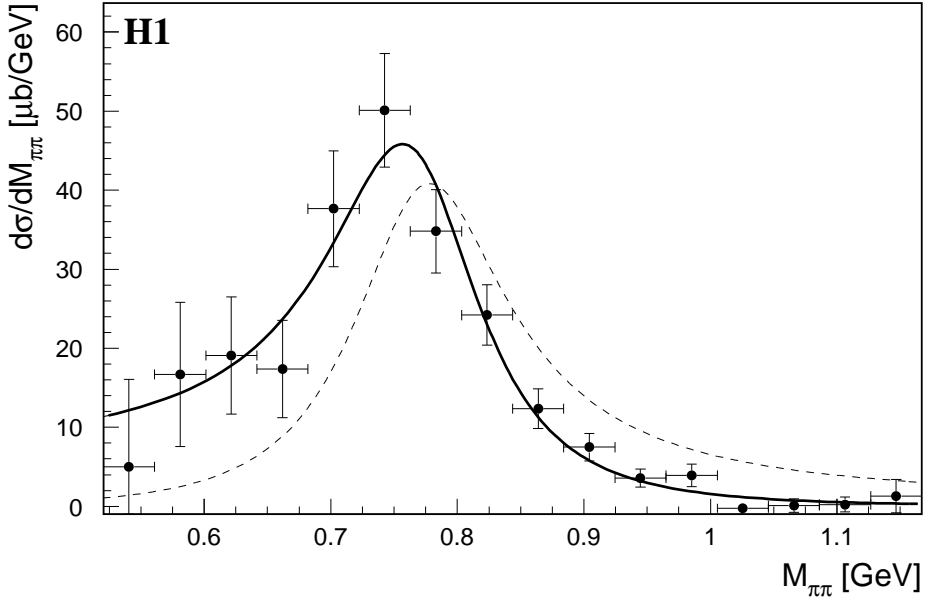,width=0.85\textwidth}
    \caption{\small The ${\rm d}\sigma/{\rm d}\mpipi$ distribution at
      $\avWgp=55$~GeV for $|t|<0.5$~${\rm GeV^2}$\@.  Only statistical errors
      are shown. The solid curve represents a fit to the distribution
      assuming the parameterisation of Ross and Stodolsky (equations~7 and
      9)\@.  The dashed curve shows the pure Breit Wigner (equation~7 with
      $n_{\rm RS}=0$).}
    \label{rhoshape} \end{center}
\end{figure}

The acceptance, $\varepsilon_{\rm bin}$, varied between $2\, \%$ and $37\,
\%$ with an average value of $15\,\%$\@.  The integrated photon flux is
$\Phi_I = 0.0621$\@.  The production cross section is shown in
figure~\ref{rhoshape}.
The mass dependence of ${\rm d}\sigma/{\rm d}\mpipi$ is consistent with a
large contribution from $\rho^0$ photoproduction. The peak is shifted to
lower values than the nominal $\rho^0$ mass and there is an enhancement
(suppression) of the cross section at values lower (higher) than the nominal
$\rho^0$ mass.  This asymmetric shape, or skew, of the distribution is a well
known feature of $\rho^0$ photoproduction which can be  attributed to an
interference between the resonant and non-resonant production of two pions.
In the phenomenological approach by Ross and Stodolsky~\cite{RS} the
differential cross section is assumed to follow
\begin{eqnarray}
  \frac{{\rm d}\sigma}{{\rm d}\mpipi}=f_\rho\cdot{\it BW}_\rho(\mpipi) \cdot
  (M_\rho/\mpipi)^{n_{\rm RS}}
  \label{eqn:rs}
\end{eqnarray}
where $f_\rho$ and $n_{\rm RS}$ are constants and
\begin{eqnarray}
  {\it BW}_\rho(\mpipi)=\frac{\mpipi M_\rho \Gamma_\rho(\mpipi)}{(M_\rho^2
    -\mpipi^2)^2+M_\rho^2\Gamma^2_\rho(\mpipi)},
  \label{eqn:bw}
\end{eqnarray}
where $M_\rho$ is the $\rho^0$ mass and where the momentum dependent width,
following the suggestion of Jackson~\cite{jackson}, is taken to be
\begin{eqnarray}
\Gamma_\rho(\mpipi)=\Gamma_0(\frac{q^*}{q^*_0})^3\frac{2}{1+(q^*/q^*_0)^2}.
\label{eqn:width}
\end{eqnarray}
Here $\Gamma_0$ is the $\rho^0$ width, $q^*$ is the $\pi$ momentum in the
$\pi^+\pi^-$ rest frame and $q^*_0$ is the value of $q^*$ when
$\mpipi=M_\rho$\@.  The distortion of the mass spectrum is characterised by
the Ross-Stodolsky parameter $n_{\rm RS}$, which becomes zero for production
of just the resonance.

The spectrum was fitted, using the parameterisation of equation~\ref{eqn:rs},
over the range $0.52<\mpipi<1.17$~GeV with quantities $f_\rho$, $M_\rho$,
$\Gamma_0$ and $n_{\rm RS}$ left as free parameters. The fit is shown in
figure~\ref{rhoshape} and the results are summarised in
table~\ref{table:rhofit}\@.  The values of $M_\rho$ and $\Gamma_0$ are in
agreement with particle data group (PDG) values of $M_\rho=769.9\pm 0.8$~MeV
and $\Gamma_0=151.2 \pm 1.2$~MeV~\cite{PDG}\@.  The value of the differential
cross section at the nominal $\rho^0$ mass was measured to be
\begin{table}[bt]
  \begin{center}
    \begin{tabular}{||l|rcll||} \hline
    Parameter      & Value  &     &      &               \\ \hline
    $M_\rho$       & 783    &$\pm$& 13   & MeV           \\ 
    $\Gamma_0$     & 153    &$\pm$& 19   & MeV           \\ 
    $f_\rho$       & 6.23   &$\pm$& 0.62 & ${\rm \mu b}$ \\ 
    $n_{RS}$       & 5.84   &$\pm$& 1.05 &               \\ 
    $\chi^2/{\rm ndf}$ & \multicolumn{3}{c}{$12.81/12$} & \\ \hline
   \end{tabular}
 \end{center}
 \caption{\small Results of the fit of the ${\rm d} \sigma/{\rm d}
\mpipi$ distribution for $\avWgp = 55$~GeV and $|t|<0.5$~${\rm GeV^2}$
assuming a Ross-Stodolsky skewing factor (see text). The errors shown
are statistical.}
 \label{table:rhofit}
\end{table}

\begin{eqnarray*}
  \left. \frac{{\rm d}\sigma}{{\rm d}\mpipi} \right|_{\mpipi=M_\rho} =44\pm
    5\,(\stat)\pm12\,(\syst)\;\mu\mbox{b}\, {\rm GeV}^{-1}.
\end{eqnarray*}
The systematic error was determined in a similar way to that described in
section~\ref{sec:crosssect}\@.  Spital and Yennie have suggested that the
value of the differential cross section at the nominal $\rho^0$ mass offers a
way to determine the total elastic $\rho^0$ cross section independently of
the assumptions made for the form of the line shape since it is at the
nominal $\rho^0$ mass where there is little or no interference of any
additional coherent production~\cite{spital&yennie}\@.  With their assumption
of a standard Breit Wigner for the $\rho^0$ resonance line shape, we get
\begin{eqnarray*}
  \sigmarho= \frac{\pi \Gamma_0}{2} \left. \frac{{\rm d}\sigma}{{\rm
      d}\mpipi} \right|_{\mpipi=M_\rho} = 10.6 \pm 1.1 (\stat) \pm 3.0
  (\syst)\;\mu\mbox{b}  
\end{eqnarray*}

In order to investigate the sensitivity of the results to changes in the
assumed parameterisation of the mass spectrum, the data were reanalysed with
a fit based on the S\"{o}ding interference model~\cite{SODING} and with
different assumptions for the momentum dependent width.  In the S\"{o}ding
model the skewing of the mass spectrum is explained by the interference of a
resonant $\rho^0\rightarrow\pi^+\pi^-$ amplitude and a p-wave $\pi\pi$ Drell
type background term~\cite{DRELL}\@.  A simplified parameterisation of the
model used in~\cite{ZEUS2,aston} can be written as
\begin{eqnarray}
  \frac{{\rm d}\sigma}{{\rm d}\mpipi}=f_\rho\cdot{\it
    BW}_\rho(\mpipi)+f_I\cdot I(\mpipi),
\label{eqn:soding}
\end{eqnarray}
where ${\it BW}_\rho(\mpipi)$ is defined in equation~\ref{eqn:bw} and the
interference term is
\begin{eqnarray}
  I(\mpipi)=\frac{M^2_\rho-M^2_{\pi\pi}}{(M_\rho^2
    -\mpipi^2)^2+M_\rho^2\Gamma^2_\rho(\mpipi)}.
\end{eqnarray}
The quantities $f_\rho$, $f_I$, $M_\rho$ and $\Gamma_0$ are left as free
parameters.  The fit was also done for both models with the following two
parameterisations of the momentum dependent width, used by some other
experiments~\cite{aston,gladding, mcclellon}
\begin{eqnarray}
  \Gamma_\rho(\mpipi)=\Gamma_0(\frac{q^*}{q^*_0})^3, \label{eqn:width1} \\ 
  \Gamma_\rho(\mpipi)=\Gamma_0(\frac{q^*}{q^*_0})^3\frac{M_{\rho}}{\mpipi}
\label{eqn:width2}
\end{eqnarray}
The results of the fits are summarised in table~\ref{tab:fits}\@.  All of the
fits yielded values for $M_\rho$ and $\Gamma_0$ in agreement apart from that
using the S\"{o}ding model with the width of equation~\ref{eqn:width1}\@.

\begin{table}[hbt]
  \begin{center}
    \begin{tabular}{||l|c|c|c|c|c|c||} \hline
      Model & $\chi^2/{\rm ndf}$& $M_\rho$ /MeV & $\Gamma_0$ /MeV& $f_\rho$ /$\mu$b & $n_{RS}$ &
       $f_I$ $/{\rm \mu b GeV^{-1}}$ \\ \hline
    RS +(8)  & 1.07 &$783 \pm 13$& $153 \pm 19$ & $6.23 \pm 0.62$ & $5.84 \pm 1.05$ & ---\\
    RS +(11) & 1.09 &$791 \pm 14$& $163 \pm 23$ & $6.11 \pm 0.65$ & $6.84 \pm 1.00$ & ---\\
    RS +(12) & 1.09 &$784 \pm 13$& $153 \pm 19$ & $6.20 \pm 0.62$ & $5.95 \pm 1.01$ & ---\\

    S +(8)   & 0.84 &$787 \pm 12$& $176 \pm 28$ & $6.70 \pm 0.55$ & --- & $3.39 \pm 0.85$ \\
    S +(11)  & 1.02 &$815 \pm 16$& $205 \pm 40$ & $5.86 \pm 0.49$ & --- & $5.07 \pm 1.20$ \\
    S +(12)  & 0.88 &$793 \pm 13$& $181 \pm 30$ & $6.57 \pm 0.54$ & --- & $3.74 \pm 0.92$ \\
    \hline
     
   \end{tabular}
 \end{center}
 \caption{\small The results of the fits of the ${\rm d}\sigma/{\rm
     d}\mpipi$ spectrum for the low $\Wgp$ analysis.  In the column
   marked `Model': RS is the Ross-Stodolsky parameterisation of
   equation~6, S is the S\"{o}ding parameterisation of equation~9\@.
   The bracketed numbers refer to the equation describing the momentum
   dependent width.  The errors quoted are only statistical.  The
   column labels are specified in the text.}
 \label{tab:fits}
\end{table}

\subsection{The  Elastic {\boldmath  $\rho^0$} Photoproduction Cross Section}
\label{sec:crosssect}

The $\gamma p \rightarrow \pi^+ \pi^- p$ cross section at $\avWgp = 55$~GeV
was obtained by integrating the fitted form of the differential invariant
mass spectrum of equation~\ref{eqn:rs} with a momentum dependent width as in
equation~\ref{eqn:width}\@.  This procedure involves an extrapolation outside
the measured region, $0.52<\mpipi<1.17$~GeV, to the region
$2M_\pi<\mpipi<M_{\rm cut}$, where, following the procedure used
by~\cite{ZEUS2}, the upper mass limit is taken to be $M_{\rm
  cut}=M_{\rho}+5\Gamma_0$\@.  In making the extrapolation we assume a $t$
dependence with a slope parameter $b=10.9$~${\rm GeV^{-2}}$, as obtained from
the fit of the ${\rm d}\sigma/{\rm d}t$ distribution in
section~\ref{sec:tfit}, so that
\begin{eqnarray}
  \frac{{\rm d}^2\sigma(\gamma p \rightarrow \pi^+ \pi^- p)}{{\rm d}\mpipi \,
    {\rm d}t}=A(\mpipi)\cdot e^{bt}.
\end{eqnarray}
The integration is performed over the limits $-0.5$~${\rm GeV^2}<t<t_{\rm
  min}$, where $|t_{\rm min}|$ is the minimum value of $|t|$ which is
kinematically allowed.  The fraction of the cross section that lies outside
the measured kinematic region in $\mpipi$ , $\xi$, is found to be
$\xi=0.15$\@.  The resulting total $\gamma p \rightarrow \pi^+ \pi^- p$ cross
section is
\begin{eqnarray*}
  \sigmapi  & = & 11.2 \pm 1.1\,(\stat)\, 
                       \pm 3.1\,(\syst)\;\mu\mbox{b}.
\end{eqnarray*}
The resonant part of the total cross section is obtained by integrating the
function \linebreak $f_\rho\cdot{\it BW}_\rho(\mpipi)$ (i.e.\ taking $n_{\rm
  RS}=0$ in equation~\ref{eqn:rs}), assuming the same $t$ dependence as
before, over the prescribed range in $\mpipi$\@.  The value of the cross
section, measured at $\avWgp=55$~GeV for $2M_\pi<\mpipi<M_\rho+5\Gamma_0$ and
$-0.5$~${\rm GeV^2}<t<t_{\rm min}$ is then
\begin{eqnarray*}
  \sigmarho  & = & 9.1 \pm 0.9\,(\stat)\, 
                       \pm 2.5\,(\syst)\;\mu\mbox{b}.
\end{eqnarray*}

At the present level of experimental precision there are no corrections to be
made for $\rho^0$ decay modes other than to $\pi^+ \pi^-$\@.

For the chosen value of $M_{\rm cut}=M_{\rho}+5\Gamma_0$ the extracted cross
section, $\sigmarho$, was found to be very stable against changes in the
parameterisation of the ${\rm d}\sigma/{\rm d}\mpipi$ spectrum.  Allowing
$n_{\rm RS}$ to vary within the statistical errors resulted in a change of
$\pm 6.0 \,\%$\@.  Repeating the fit with values for $M_\rho$ and $\Gamma_0$
fixed to the PDG values gave an increase of $7.0\,\%$\@.  Using the
parameterisation of equation~\ref{eqn:soding} gave an increase of the cross
section of $5.9\,\%$ and the maximum variation caused by different
assumptions of the momentum dependence of the width was $6.6\,\%$\@.

If, however, we do not impose an upper $M_{\pi \pi}$ limit but use the
parameterisations to extrapolate over the full kinematic range available
(i.e.\ $M_{\rm cut}=\Wgp-M_p$) as was the procedure for measurements made at
lower energies, we find a very large variation in the cross section extracted
with the various methods (see table~\ref{tab:fitsXS} for a summary of all the
fits)\@.  This is because some of the parameterisations assume the $\rho^0$
resonance to have a long tail extending up to high values of $\mpipi$\@. 
Unlike the situation at lower energies kinematic constraints do not suppress
this tail to levels where it can be neglected.

\begin{table}[hbt]
  \begin{center}
    \begin{tabular}{||l|c|c|c|c||} \hline
      Model & \multicolumn{2}{|c|}{$M_{cut}=M_{\rho}+5\Gamma_0$} 
          & \multicolumn{2}{|c||}{$M_{cut}=\Wgp-M_p$} \\ \hline
        &   $\sigma_{\pi\pi}$\,$/ \mu\mbox{b}$ & $\sigma_\rho$\,$/ \mu\mbox{b}$
        &   $\sigma_{\pi\pi}$\,$/ \mu\mbox{b}$ & $\sigma_\rho$\,$/ \mu\mbox{b}$ \\ \hline
    RS +(8)  &  11.2 &  9.1 & 11.2 & 10.5 \\
    RS +(11) &  11.4 &  9.3 & 11.4 & 15.9 \\
    RS +(12) &  11.5 &  9.2 & 11.5 & 13.5 \\ 
                                                                                   \hline
    S +(8)   &  11.2 & 9.6 &  10.8 & 11.4 \\
    S +(11)  &  12.3 & 8.5 &  17.0 & 14.5 \\
    S +(12)  &  11.5 & 9.6 &  14.5 & 14.8 \\
    \hline
     
   \end{tabular}
 \end{center}
 \caption{\small The extracted $\sigmapi$ and $\sigmarho$ cross
   sections from the fits of the ${\rm d}\sigma/{\rm d}\mpipi$
   spectrum for two choices of $M_{\rm cut}$ (see text) for the low
   $\Wgp$ analysis.  In the column marked `Model' RS is the
   Ross-Stodolsky parameterisation of equation~6, S is the S\"{o}ding
   parameterisation of equation~9\@.  The bracketed numbers refer to
   the equation describing the momentum dependent width.  }
 \label{tab:fitsXS}
\end{table}

Since it was not possible to determine the $\mpipi$ and $t$ dependence of the
cross section at $\avWgp = 187$~GeV, we took the values of $n_{\rm RS}$ and
$b$ found in the low $\Wgp$ analysis\footnote{This procedure is likely to be
  valid since the measured value of $n_{\rm RS}$ is close to values obtained
  at lower energies~\cite{aston, gladding, eisenberg}, indicating there is
  little dependence of $n_{\rm RS}$ on $\Wgp$.} to reweight the MC
distributions used for the acceptance calculation and to determine
numerically $\xi$ (the fraction of the cross section outside the measured
region) and the ratio of the cross sections $r_{\rm
  nr}=\sigmarho/\sigmapi$\@.  The values obtained were $\xi=0.05$ and $r_{\rm
  nr}=0.82$\@.  The elastic $\rho^0$ cross section was then determined using
\begin{eqnarray}
  \sigmarho =\frac{N-N_{\rm bgd}}{\Phi_I {\cal L} \varepsilon} \cdot 
  {1\over 1-\xi} \cdot r_{\rm nr},
\end{eqnarray}
where $N$ is the measured number of events in the mass range
$0.4<\mpipi<1.26$~GeV, $N_{\rm bgd}$ is the estimated number of background
events, $\Phi_I=0.00903$ the photon flux found using equation~\ref{eq:ijray},
${\cal L}$ is the integrated luminosity and $\varepsilon$ is the acceptance.
The elastic $\gamma p \rightarrow \rho^0 p$ cross section for $-0.5$~${\rm
  GeV^2}<t<t_{\rm min}$ and $2M_\pi<\mpipi<M_{\rho}+5\Gamma_0$ at $\avWgp =
  187$~GeV is then
\begin{eqnarray*}
  \sigmarho  & = & 13.6 \pm 0.8\,(\stat)\, 
                       \pm 2.4\,(\syst)\;\mu\mbox{b}.
\end{eqnarray*}
 
The breakdown of the various contributions to the systematic error of the
elastic $\rho^0$ cross section measurement at $\avWgp=55$~GeV for the mass
range $2M_\pi<\mpipi<M_{\rho}+5\Gamma_0$ can be found in
table~\ref{table:systematic}\@.  The primary source of systematic error comes
from the uncertainty of the trigger efficiency, where the large error of
$20\, \%$ is due to the low statistics in the overlap sample of events used
to determine it.  An error of $5\, \%$ is assigned to the uncertainty in the
scanning yield, which was estimated by taking the spread in the yield as
found by several scanners. The track fitting error of $4\, \%$ was estimated
by taking the difference between the acceptances found in data and MC\@.  An
error of $3\, \%$, arising mainly from limited statistics, was assigned to
the MC acceptance. Changing the $t$ dependence within the errors of the
measurement gave a $12\, \%$ error.  To estimate the error due to background
contributions each background was varied by $\pm 50\, \%$ and the $\mpipi$
distribution re-fitted. An error of $6\, \%$ was assigned to reflect the
spread in the extracted cross section due to different assumptions for the
parameterisation of the $\mpipi$ dependence. The error in the integrated
luminosity measurement was $5\,\%$\@.

\begin{table}[hbt]
  \begin{center}
    \begin{tabular}{||l|r||} \hline
      Source                & Error    \\ \hline \hline 
      Trigger efficiency    & $20\,\%$ \\ \hline
      Scanning yield        & $ 5\,\%$ \\ \hline
      Track fit efficiency  & $ 4\,\%$ \\ \hline
      MC acceptance         & $ 3\,\%$ \\ \hline
      GD background         & $ 8\,\%$ \\ \hline
      PD background         & $ 5\,\%$ \\ \hline
      DD background         & $ 2\,\%$ \\ \hline
      ND background         & $ 1\,\%$ \\ \hline
      EL background         & $ 1\,\%$ \\ \hline
      $t$ dependence        & $12\,\%$ \\ \hline
      Resonance extraction  & $ 6\,\%$ \\ \hline 
      Luminosity            & $ 5\,\%$ \\ \hline \hline
      Total systematic  error    & $28\,\%$ \\ \hline
    \end{tabular}
  \end{center}
  \caption[Systematic errors for cross section]{\small Sources of systematic
    errors for the elastic \r0 cross section measurement at $\avWgp =
    55$~GeV.}
  \label{table:systematic}
\end{table}

For the high $\Wgp$ analysis the biggest uncertainty on the elastic $\rho^0$
cross section measurement comes from background in the elastic $\rho^0$
sample.  A systematic error was taken of $50\, \%$ of each background
estimated using the MC simulations and $25\, \%$ for the electron beam gas
background. The error on the electron tagger efficiency was $5\,
\%$~\cite{NewH1}\@.  An error of $6\, \%$ was assigned for the MC acceptance
being due to the difference between PHOJET and PYTHIA and the MC statistics.
An error of $1\, \%$ was found by varying the $t$ slope parameter by $\pm
4$~${\rm GeV^{-2}}$ around the central value of $11$~${\rm GeV^{-2}}$\@.  The
$20\, \%$ overall hadronic energy scale uncertainty of the BEMC~\cite{H1f2}
gave rise to a $4\, \%$ error. A $6\, \%$ error was assigned to account for
the uncertainty in the method of extraction of the resonant cross section.
This error was estimated by taking the spread in the value of the cross
section obtained using the results of each fit to the low $\Wgp$ data as
described above, reweighting the input MC $\mpipi$ distributions and
recalculating the acceptances and values for $\xi$ and $r_{\rm nr}$\@.  The
error in the integrated luminosity measurement was $5\,\%$\@.  A breakdown of
the systematic error is shown in table~\ref{table:systematicII}\@.

\begin{table}[p]
\begin{center}
  \begin{tabular}{||l|r||} \hline
    Source               & Error     \\ \hline \hline
    GD background        & $9\,\%$    \\ \hline
    PD background        & $4\,\%$    \\ \hline
    DD background        & $2\,\%$    \\ \hline
    ND background        & $1\,\%$    \\ \hline
    EL background        & $7\,\%$    \\ \hline
    MC acceptance        & $6\,\%$    \\ \hline
    BEMC hadronic scale  & $4\,\%$    \\ \hline
    $t$ dependence       & $1\,\%$    \\ \hline
    e-gas background     & $5\,\%$    \\ \hline
    e-tagger acceptance  & $5\,\%$    \\ \hline 
    Resonance extraction & $6\,\%$    \\ \hline 
    Luminosity           & $5\,\%$    \\ \hline \hline
    Total systematic error    & $18\,\%$
    \\ \hline
  \end{tabular}
\end{center}
 \caption[Systematic errors for cross section]{\small Sources of systematic
   errors for the elastic \r0 cross section measurement at
   $\avWgp = 187$~GeV.} 
 \label{table:systematicII}
\end{table} 

\begin{figure}[bp]
  \vspace*{-1.5cm}
 \begin{center} \epsfig{file=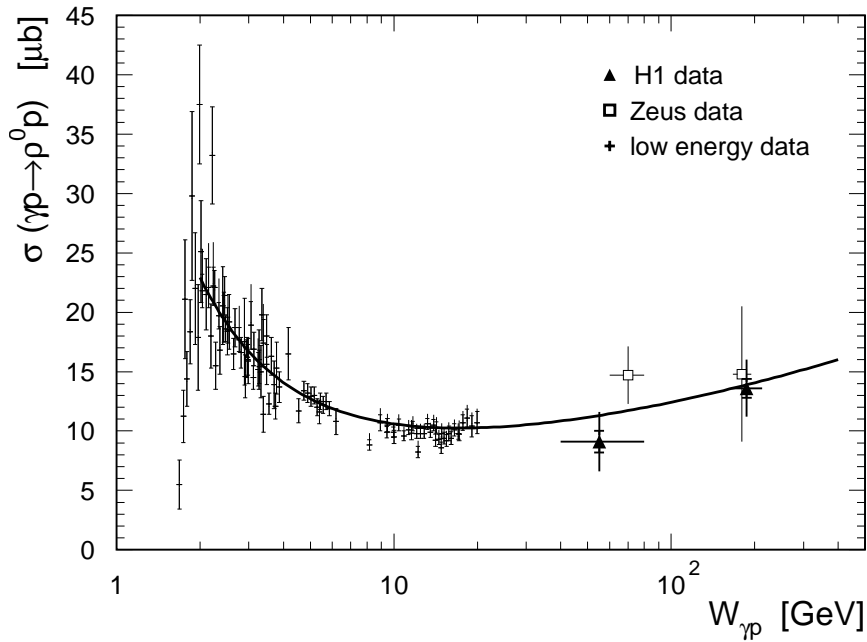,width=0.8\textwidth}
\caption[tot  cross section]
{\small The dependence of the elastic $\rho^0$ photoproduction cross section,
  $\sigma (\gamma p \to \rho ^0 p)$, on $W_{\gamma p}$ for H1 and previous
  measurements~\cite{ZEUS1,ZEUS2,baldini,callahan}\@.  The inner vertical bars
  on the H1 points denote only the statistical error, the outer ones contain
  statistical and systematic errors added in quadrature. The horizontal error
  bars show the range in $\Wgp$ over which the measurements were made. The
  solid curve shown is a curve based on a soft pomeron calculation, the
  normalisation of which is fixed by the data at low
  energy~\cite{SchSjoNUCB}.}
   \label{fig:xsect} 
 \end{center}
\end{figure}

\pagebreak

Figure~\ref{fig:xsect} summarises the dependence on $\Wgp$ of measurements of
the elastic \r0 photoproduction cross
section~\cite{ZEUS1,ZEUS2,baldini,callahan} including our new measurements.
Also shown is the theoretical prediction based on the VMD model and
non-perturbative Regge theory~\cite{SchSjoNUCB} (solid curve).  The new high
energy measurements at HERA demonstrate that the $\Wgp$ dependence of the
cross section is consistent with the slight increase characteristic of this
model in which the leading exchange is a `soft' pomeron.

\subsection{The {\boldmath $t$} Dependence of the Elastic {\boldmath $\rho^0$}
  Photoproduction Cross Section}
\label{sec:tfit}

Diffractive scattering is characterised by a low momentum transfer between
the vector meson and the proton, resulting in a steeply falling distribution
in $t$\@.  This is quantified by the slope parameter, $b$, determined by
fitting the form $ e^{bt}$ to the $t$ dependence of ${\rm d}\sigma/{\rm d}t$
for the data taken at $\avWgp=55$~GeV\@.

\begin{figure}[hb]
  \vspace*{-0.5cm}
  \epsfysize=9truecm 
  \centerline{\epsfbox{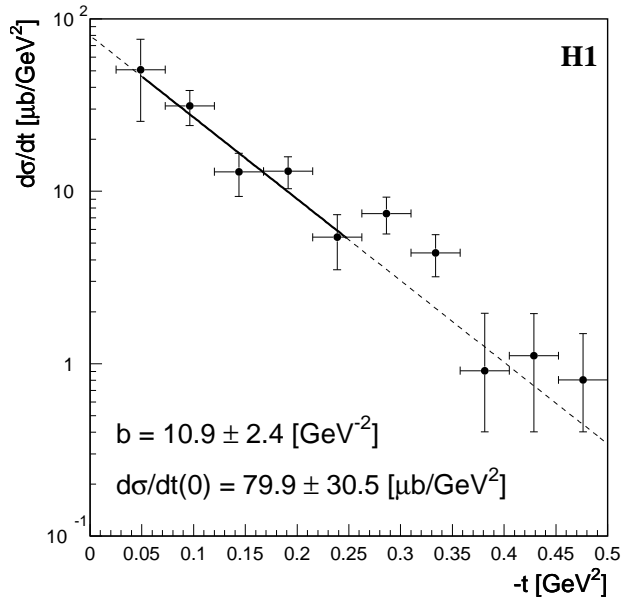}} 
  \caption {\small The differential cross section, ${\rm d}\sigma/{\rm d}t$,
    at $\avWgp=55$~GeV is fitted over the range $0.025<|t|<0.25$~GeV$^2$ with
    the function ${\rm d}\sigma/{\rm d}t=a\, e^{bt}$ (only statistical errors
    are shown).}
  \label{fig:tfit}
\end{figure}

The cross section ${\rm d}\sigma/{\rm d}t$ was determined by integrating the
mass spectrum over the measured range ($0.52<M_{\pi \pi}<1.17$~GeV) and
correcting for losses and backgrounds using the MC simulation. A correction
was made for losses in the tails making the assumption that $n_{\rm RS}$, the
Ross-Stodolsky parameter, remained constant with $t$, so that $\xi=0.15$ at
each value of $t$\@.  The results are shown in figure~\ref{fig:tfit}\@.  The
data for $0.025<|t|<0.25$~GeV$^2$ are well parameterised with a simple
exponential with slope parameter
\begin{eqnarray*}
  b=10.9 \pm 2.4\,(\stat) \pm 1.1\,(\syst) \;{\rm GeV^{-2}}
\end{eqnarray*}
and intercept 
\begin{eqnarray*}
  \left . {{\rm d}\sigmarho \over {\rm d}t}\right|_{t=0}
  = 79.9 \pm 30.5 \,(\stat) \pm 21.9\,(\syst) \;{\rm \mu b/GeV^2}.
\end{eqnarray*}
The systematic error arises from the variation of the MC input parameters for
the skewing of the line shape and the $b$ slope within the statistical errors
of the fits and also by allowing each background to vary by $50\,\%$\@.  For
the intercept the systematic errors on the normalisation are also taken into
account (see section~\ref{sec:crosssect}).  For $|t|>0.25$~${\rm GeV}^2$,
deviations from a simple exponential dependence were shown by other
measurements of diffractive vector meson photoproduction~\cite{aston,
  gladding, barber}\@.

\begin{figure}[htbp]
  \vspace*{-1.0cm}
  \epsfysize=3.2truein
  \centerline{\epsfbox{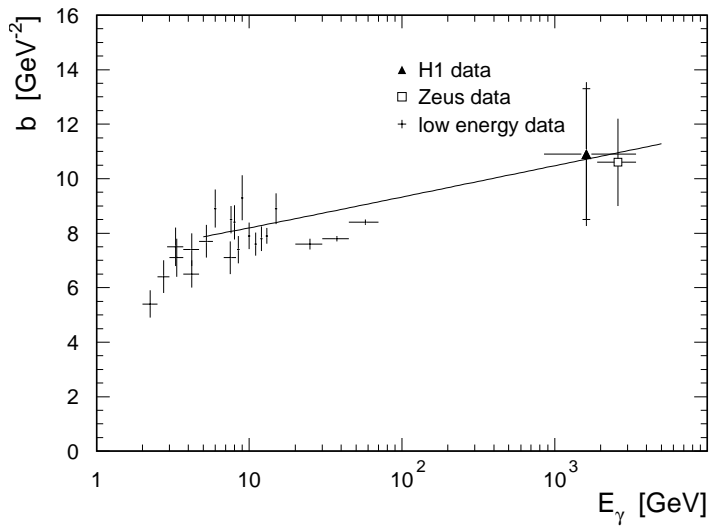}}
  \caption[Comparison of $b_t$ measurement with previous data.]%
  {\small Determinations of the exponential slope parameter, $b$, as a
    function of $E_{\gamma}$, the photon energy in the rest frame of the
    proton.  The inner vertical bar on the H1 point denotes the
    statistical error only, the outer one contains statistical and systematic
    errors added in quadrature. The horizontal error bars show the range in
    $\Wgp$ over which the measurements were made.  The curve is a prediction
    based on pomeron exchange~\cite{SchSjoNUCB}.}
  \label{fig:bt}
\end{figure}

The measurement of $b$ is compared to previous determinations in
figure~\ref{fig:bt}\@.  The recent ZEUS measurement~\cite{ZEUS2} is shown
along with lower energy data as compiled by Aston et al.~\cite{aston,
  gladding, mcclellon, eisenberg, berger}\@.  The ZEUS and H1 results
together show that the shrinkage of the $t$ dependence of ${\rm d}\sigma/{\rm
  d}t$ continues into the HERA energy range as expected in the soft pomeron
picture.

\begin{figure}[hbtp]
  \vspace*{-1.0cm}
  \begin{center}
    \mbox{\epsfysize=8.0cm \epsffile{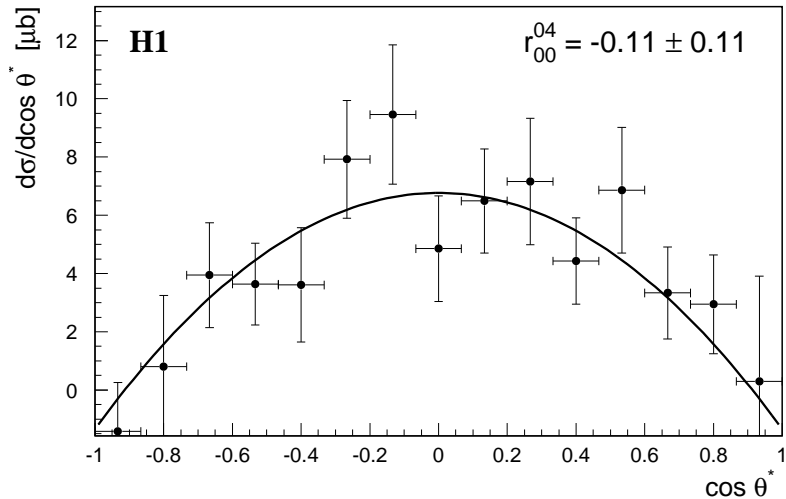}} 
  \end{center}
  \caption{\small The acceptance corrected  $\cos\theta ^*$ density
    distribution at $\avWgp=55$~GeV\@.  The data are fitted to the function
    in equation~16\@.  The errors quoted are only statistical.}
  \label{fig:costta}
\end{figure}

\subsection{Polar Angular Decay Distribution of the {\boldmath $\rho^0$}}

A feature of the diffractive production of hadrons, in addition to a slight
dependence of the cross section on $\Wgp$ and a peripheral $t$ dependence, is
s-channel helicity conservation.  This is best studied in the helicity frame,
defined as the rho meson rest frame, with the quantisation axis along the
direction of the rho in the $\gamma p$ frame~\cite{jackson}\@.  In this
frame, the polar angle distribution of the $\pi^+$ according to the formalism
in~\cite{shilling,BSYP78} is
\begin{equation}
  \frac{{\rm d}\sigma}{{\rm d}\cos \theta ^*} \propto (3\rr-1) \cos^2 \theta
  ^* + (1-\rr),
  \label{eq:wcos}
\end{equation}
where $\rr$ is the spin density matrix element which specifies the
probability that the $\rho^0$ meson is produced in the spin substate 0\@.

In order to determine $\rr$, the distribution in $\cos \theta ^*$ was
obtained using the same selection, subtraction and correction procedures as
for the ${\rm d}\sigma/{\rm d}t$ analysis. The expression~\ref{eq:wcos} was
then fitted to this distribution to obtain a value of
\begin{eqnarray*}
\rr = -0.11 \pm 0.11\,(\stat) \pm 0.04\,(\syst).
\end{eqnarray*}
Figure~\ref{fig:costta} shows ${\rm d}\sigma/{\rm d}\cos \theta ^*$ and the
fit of the expression~\ref{eq:wcos}\@.  In the helicity frame the $\rho^0$ is
thus observed to be produced predominantly in spin substates $\pm 1$ and not
in spin substate 0, consistent with the hypothesis of s-channel helicity
conservation at $Q^2=0$.


\section{Conclusions}

With the H1 detector we have measured the cross section for the elastic
photoproduction of \r0 mesons and found $\sigmarho = 9.1\pm 0.9\,(\stat)\pm
2.5\,(\syst)\;\mu\mbox{b}$ at $\avWgp = 55$~GeV\ and $\sigmarho = 13.6\pm
0.8\,(\stat)\pm 2.4\,(\syst)\;\mu\mbox{b}$ at $\avWgp = 187$~GeV\@ for
$2M_\pi<\mpipi<M_{\rho}+5\Gamma_0$\@.  At the former $\Wgp$, the slope
parameter of the distribution in $t$ is $10.9 \pm 2.4\,(\stat) \pm
1.1\,(\syst) \mbox{ GeV}^{-2}$ and the decay polar angular distribution is
consistent with a pure $\sin^2\theta ^*$ distribution.  We thus verify the
extension to HERA energies of the properties found at lower energies for
elastic photoproduction of the \r0 meson, namely a steep forward distribution
of the produced \r0 mesons with a continued shrinkage of the $t$ dependence
with increasing $\Wgp$ and a decay angular distribution that is consistent
with $s$-channel helicity conservation.  A slow rise of the elastic $\rho^0$
cross section with $W_{\gamma p}$, as expected in a picture of interactions
based upon the soft pomeron which also matches the rise of the total
photoproduction cross section, is shown to be consistent with observation.


\section*{Acknowledgements}

We are grateful to the HERA machine group whose outstanding efforts made this
experiment possible. We appreciate the immense effort of the engineers and
technicians who constructed and maintained the detector.  We thank the
funding agencies for their financial support of the experiment. We wish to
thank the DESY directorate for the support and hospitality extended to the
non-DESY members of the collaboration.


\end{document}